\documentclass[ reprint,
 amsmath,amssymb,
 aps,
]{revtex4-1}
\usepackage{caption}
\usepackage{graphicx}
\usepackage{dcolumn}
\usepackage{bm}

\begin{document}

\preprint{APS/123-QED}

\title{Embedded Energy State in an Open Semiconductor Heterostructure }

\author{Ivana Hrebikova}
\email{hrebiiva@fel.cvut.cz}

\author{Lukas Jelinek}%
\email{lukas.jelinek@fel.cvut.cz}
\affiliation{%
 Department of Electromagnetic Field, Faculty of Electrical Engineering, Czech Technical University in Prague, 166 27 Prague, \mbox{Czech republic}
}%

\author{M\'ario G. Silveirinha}%
\email{mario.silveirinha@co.it.pt}
\affiliation{%
	Department of Electrical Engineering, Instituto de Telecomunica\c{c}\~{o}es, University of Coimbra, 3030-290 Coimbra, Portugal
}%



\date{\today}

\begin{abstract}
In this paper, we show that HgCdTe heterostructures may support bound electronic states embedded in the continuum, such that the discrete energy spectrum overlaps the continuous spectrum. Although the proposed heterostructures are generally penetrable by an incoming electron wave, it is shown that they may support spatially localized trapped stationary states with an infinite lifetime. We discuss the possibility of a free electron being captured by the proposed open resonator, and present a detailed study of the trapping lifetime in the case of a detuned resonator.
\begin{description}
\item[PACS numbers]
{73.21.-b}, {73.23.Ad}, {73.63.Kv}, {41.20.-q}
\end{description}
\end{abstract}

\maketitle


\section{\label{sec:level1}Introduction}

The stationary states of a quantum system with a finite height potential well are commonly divided into bound states, which form a discrete spectrum, and unbound states which form a continuum \cite{Gottfried-2003}. Usually, the two classes of modes do not overlap: the energies of the bound states usually lie within a potential well, while the energies of the unbound states lie above the potential well. Surprisingly, this property of a quantum system is not universal, as there are theoretical predictions of systems with bound state energies falling into the continuum, so-called bound states in the continuum - BICs. Pioneering work pointing out that bound states with energies in the continuum are exact solutions of the one-electron Schr\"odinger equation for specific potentials was presented by von Neumann and Wigner in 1929 \cite{Neumann-1929}. This paper also showed how to design electric potentials supporting BICs. The original formulation of von Neumann and Wigner has been reworked and even extended to a two-electron wave function \cite{Stillinger-1975}, still bearing the sign of BICs.

Alongside the paradigm introduced above, so-called ``resonant states" in quantum systems have been discovered \cite{Feshbach-1958}. These represent a different approach for achieving ``bound" states (resonances) with energies lying above the continuum threshold. These narrow--width resonances were proposed to exist as metastable states trapped by a large potential barrier, or as quasi--bound states in closed channels of a system with weakly coupled channels \cite{Newton, Bulgakov-2006, Sadreev-2006, Moiseyev-2009}. Strictly speaking, however, these states are not truly bound, as they are in fact localized states with a finite lifetime constructed from continuum states. Their appearance is however very close to true BICs. Up to now, the emergence of BICs in quantum systems has never been demonstrated experimentally. The experiment closest to observing BICs was carried out by Capasso in 1992 \cite{Capasso-1992}. The ``bound state", albeit with energy above the potential barrier, was $de facto$ a defect mode achieved by Bragg reflections in the periodic system of potential wells \cite{Weber-1994}.

The resonant states in the continuum have recently elicited significant interest in the field of photonics. Indeed, for light waves it may be easier to design a resonator environment at will, using photonic crystals or metamaterials \cite{Gippius-2005, Borisov-2005, Marinica-2008, Silveirinha-2014a, Alu-2014}. Photonic crystals even made an experimental observation of BICs possible \cite{Plotnik-2011, Lee-2012, Hsu-2013}. Until recently, all the known realizations of BIC resonators required infinitely extended material profiles, e.g. a photonic crystal. Truncation of the material profile leads to imperfect localization and to finite oscillation lifetimes. Importantly, it was shown for the first time in \cite{Silveirinha-2014a} that spatially unbounded resonators are not required to have BICs, and that, under some strict conditions, volume plasmons may enable the formation of BICs in open cavities of finite size.

In this paper we propose a semiconductor heterostructure supporting BICs, although it is characterized by a potential well of finite height. Inspired by \cite{Silveirinha-2014a} and using an electron--light wave analogy, we show that an electron can be trapped with an infinite lifetime within a spherical core--shell heterostructure when the electron dispersive mass in the shell is precisely zero and the radius of the core is precisely tuned.
\section{\label{sec:level2}A Core--shell Trap For Electrons}
\subsection{\label{subsec:A}Electron--Light Analogy}
The trapped electron states in a crystalline heterostructure are determined by the microscopic time--independent Schr\"odinger equation
\begin{equation}
- \frac {\hbar^2}{2m_{\mathrm{e}}}\Delta \psi _{\mathrm{mic}}\left( {\bf{r}} \right) + V_{\mathrm{mic}}\left( {\bf{r}} \right)\psi _{\mathrm{mic}}\left( {\bf{r}} \right) = E\psi _{\mathrm{mic}}\left( {\bf{r}} \right)
\label{eq1}
\end{equation}
where $\Delta$ is the Laplacian operator, $\psi_\mathrm{mic} \left( \bf{r} \right)$ is the wave function of an electron from the top energy shells, $E$ is energy, $V_\mathrm{mic} \left( \bf{r} \right)$ is the effective microscopic potential associated with the ion lattice of period $a$ and with the rest of the electrons \cite{Harald}.

In realistic heterostructures, where each layer is composed of many atoms and the  wave vectors of the electron are small  $k \ll 2\pi /a$, the description of the valence electrons can however be further simplified. In such a case, it is possible to homogenize the microscopic wave function $\psi_\mathrm{mic} \left( \bf{r} \right)$ and the potential $V_\mathrm{mic} \left( \bf{r} \right)$ \cite{Kane-1957, Bastard-1986, Silveirinha-2012a}, resulting in an effective ``macroscopic" wave function $\psi _\mathrm{c} \left( \bf{r} \right) = \left\langle \psi _\mathrm{mic} \right\rangle$, which varies slowly on the scale of the lattice constant, and in an effective potential $V _\mathrm{eff} \left( \bf{r} \right) = \left\langle V _\mathrm{mic} \right\rangle $, which is a constant for each heterostructure layer \cite{Kane-1957, Bastard-1986, Silveirinha-2012a,  Lannebere-2015, Silveirinha-2012c, Silveirinha-2012b}. The brackets $\left\langle \, \right\rangle $ represent the operation of spatial averaging. This $envelope wavefunction$ formalism was originally introduced by G. Bastard \cite{Bastard-1986, Bastard, Bastard-1982}, and it was further reworked in recent studies \cite{Silveirinha-2012a, Silveirinha-2012c, Silveirinha-2012b}. The point of view of this article is based on the ideas of Ref. \cite{Silveirinha-2012a, Silveirinha-2012b}.
	
An important observation is that $\psi_{\mathrm{eff}}=\left\langle \psi_\mathrm{mic} \right\rangle$  does not imply that ${\left| \psi _{\mathrm{eff}} \left( \bf{r} \right) \right|^2} = \left\langle \left| \psi _\mathrm{mic} \right|^2 \right\rangle $, and hence $\left| \psi _{\mathrm{eff}} \left( \bf{r} \right) \right|^2$ does not generally represent the probability density. The spatially averaged probability density of energy eigenstates can be written in terms of $\psi_\mathrm{eff}$ as \cite{Silveirinha-2012a, Lannebere-2015, Fernandes-2014} (see Appendix A)
\begin{equation}
\left\langle {\left| \psi _\mathrm{mic} \right|}^2 \right\rangle  = \left( {1 - \frac{{\partial {V_{{\mathrm{eff}}}}}}{{\partial E}}} \right){\left| {{\psi _{\mathrm{eff}}}} \right|^2} + \frac{{{\hbar ^2}}}{{{m^2}}}\frac{{\partial m}}{{\partial E}}{\left\| {\nabla {\psi _{\mathrm{eff}}}} \right\|^2}.
\label{eq2}
\end{equation}
Within this paradigm, the wave function $\psi_\mathrm{eff}$ satisfies the macroscopic time--independent Schr\"odinger equation
\begin{equation}
- \frac {\hbar ^2}{2m} \Delta \psi _\mathrm{eff} \left( \bf{r} \right) + V_\mathrm{eff} \psi _\mathrm{eff} \left( \bf{r} \right) = E \psi _\mathrm{eff} \left( \bf{r} \right),
\label{eq3} 
\end{equation}
where $V_\mathrm{eff} = E_{\Gamma_6}$ is the band edge energy of the conduction ($\Gamma_6$) band, and where $m$ is dispersive effective mass \cite{Kane-1957, Bastard-1986, Silveirinha-2012a, Silveirinha-2012b}, defined as
\begin{equation}
\frac{1}{m} = \frac{1}{m_\mathrm{e}} + v_p^2\left( \frac{2}{E- E_{\Gamma_8}} + \frac{1}{E - E_{\Gamma_7}} \right).
\label{eq4}
\end{equation}
Here, $v_p^2={2P^2}/{3\hbar^2}$, and $P$ is the Kane's parameter \cite{Kane-1957}, which determines the curvature of the bands. The energy $E_{\Gamma_8}$ is the band edge energy of the valence band and the energy $E_{\Gamma_7}$ is the band edge energy of the spin--orbit split--off band. The dispersive mass depends on the electron energy, and, in general, it differs from the effective mass calculated from the curvature of the band structure.

Follows the relevant physics at the interfaces the wave function $\psi_\mathrm{eff}\left( \bf{r} \right) $ follows the relevant physics at the interfaces of the layers, equation (\ref{eq3}) is further complemented with boundary conditions, i.e. with the continuity of  $\psi_\mathrm{eff}$ and $\partial_n \psi_\mathrm{eff} / m$ at each boundary, where $\partial_n=\partial / \partial n$ and $n$ represents the direction normal to the boundary surface \cite{Bastard, Silveirinha-2012b}. For convenience, we introduce the function $\tilde {\psi} _\mathrm{eff}\left( \bf{r} \right) = \psi _\mathrm{eff}\left( \bf{r} \right)/m$, which proves useful for handling the limit $m \to 0$ ($E=E_{\Gamma_8}$), which will be discussed later. Note that the boundary conditions satisfied by $\tilde{\psi}_\mathrm{eff}$ are the continuity of $m \tilde{\psi}_\mathrm{eff}$ and the continuity of $\partial_n \tilde{\psi}_\mathrm{eff}$.

In this paper, we consider heterostructures with spherical symmetry (each heterostructure layer is a spherical shell). In such a case, one can look for a solution of Eq. (\ref{eq3}) of the form $\tilde {\psi} _\mathrm{eff} \left( \bf{r} \right) = \tilde {R}_n\left( r \right){P_n}\left( \mathrm{cos}{\theta} \right)$, where ${P_n}\left( \cos{\theta} \right)$  represents an Legendre polynomial of order $n$ \cite{Arfken}. It is straightforward to show \cite{Arfken} that the time--independent Schr\"odinger equation reduces to 
\begin{equation}
\frac{1}{r^2}\frac{\partial }{\partial r}\left[ r^2\frac{\partial \tilde {R}_n}{\partial r} \right] + \left[ \frac{2m}{\hbar^2} \left( E - E_{\Gamma_6} \right) - n\left( {n + 1} \right)\frac{1}{r^2} \right]\tilde {R}_n = 0.
\label{eq5}
\end{equation}
Due to the orthogonality of the spherical harmonics and the spherical symmetry of the system, the boundary conditions for  $\tilde {\psi}_\mathrm{eff} \left( \bf{r} \right) $ reduces to the continuity of $m \tilde {R}_n \left( r \right) $ and $\partial_r \tilde{R}_n \left( r \right) $ at each heterostructure boundary.	
\begin{figure*}[!t]
	\includegraphics{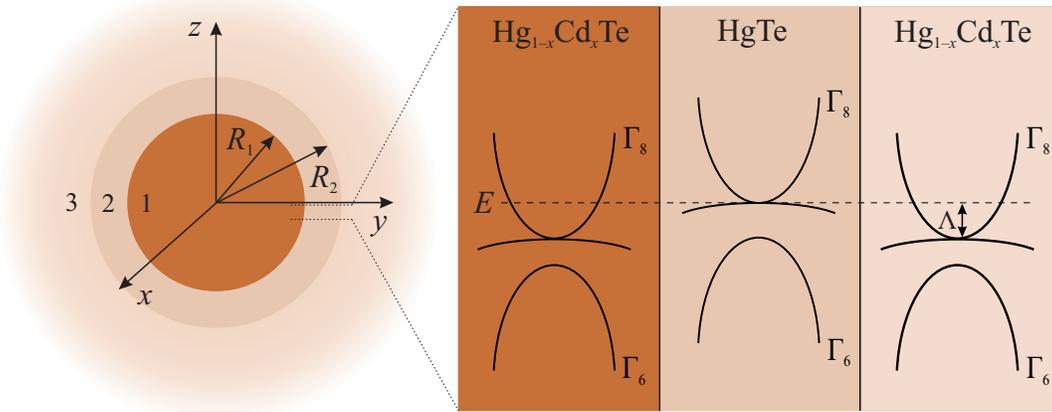}
	\caption{(color online) (left) A sketch of an open core--shell resonator for electrons consisting of a core with radius $R_1$ and a shell of width $R_2-R_1$. The core--shell structure is surrounded by an infinite background material. (right) A sketch of the energy band structures of the heterostructure. The energy level associated with the embedded energy state is represented by a dashed horizontal line, and corresponds to the edge of the valence ($\Gamma_8$) band of HgTe. Number 1 denotes the core, number 2 denotes the shell, and number 3 denotes the background.}
	\label{fig1}
\end{figure*}

Unlike an electron in a crystalline heterostructure, a light bound mode in an electromagnetic heterostructure is described by a vector wave equation. In general, the vector wave equation does not reduce to three uncoupled scalar equations, so there is no immediate analogy between the light case and the electron wave case. In the case of spherical coordinates \cite{Arfken}, the electromagnetic fields can fortunately be separated into transverse electric radial $\mathrm{TE}^r$ waves and transverse magnetic radial $\mathrm{TM}^r$ waves (transverse with respect to the radial direction) \cite{Balanis}. The $\mathrm{TE}^r$ and the $\mathrm{TM}^r$ waves can be derived from a single component of the electric vector potential $\mathrm{\bf{F}}=\mathrm{\bf{\hat{r}}}F_r$ , and the $\mathrm{TM}^r$ waves can be derived from a single component of the magnetic vector potential $\mathrm{\bf{A}}=\mathrm{\bf{\hat{r}}}A_r$, so that the vector wave equation reduces to a scalar wave equation \cite[p.553-557]{Balanis}.
\begin{subequations}
	\begin{eqnarray}
	\left( \Delta + k^2\right) \frac{F_r}{r} =0,
	\label{eq6a} \\
	\left( \Delta + k^2\right) \frac{A_r}{r} =0.
	\label{eq6b}
	\end{eqnarray}
\end{subequations}
By analogy with the electronic case, we introduce auxiliary functions  ${\tilde F_r} = {F_r}/\mu r$ and ${\tilde A_r} = {A_r}/\varepsilon r$, so that the wave equations (\ref{eq6a}) and (\ref{eq6b}) are further complemented by boundary conditions that impose the continuity of $\mu {\tilde F_r}$ and ${\partial _r}{\tilde F_r}$ for the $\mathrm{TE}^r$ waves, and the continuity of $\varepsilon {\tilde A_r}$ and ${\partial _r}{\tilde A_r}$ for the $\mathrm{TM}^r$ waves, where $\varepsilon$, $\mu$ are the permittivity and the permeability, respectively.

By comparing of (\ref{eq3}) and (\ref{eq6a}), (\ref{eq6b}) and the corresponding boundary conditions, a direct analogy between the semiconductor and electromagnetic cases is obtained, see Table \ref{table}. 
\begin{table}[!h]
	\caption{The analogy between an electron wave and electromagnetic waves.}
	\begin{tabular}{!{\vrule width 1pt}c|c|c!{\vrule width 1pt}} \noalign{\hrule height 1pt}
		\textbf{Electron wave}       									& {\textbf{TE wave}}            & {\textbf{TM wave}}\\ \hline
		${\tilde \psi _\mathrm{eff}} = {\psi _\mathrm{eff}}/m$    							& ${\tilde F_r} = {F_r}/\mu r$  & ${\tilde A_r} = {A_r}/\varepsilon r$ \\ \hline
		$m$                                 							& $\mu $                        & $\varepsilon $                       \\ \hline
		$2\left( {{E} - {E_{{\Gamma _6}}}} \right)$        & $\varepsilon $                & $\mu $                               \\ \hline 
		${k^2} = \frac{2m\left( {{E} - {E_{{\Gamma _6}}}} \right)}{\hbar ^2}$ & ${\omega ^2}\varepsilon \mu $ & ${\omega ^2}\varepsilon \mu $\\ \hline 
		\multicolumn{3}{!{\vrule width 1pt}c!{\vrule width 1pt}}{$\Delta f + {k^2}f = 0$ , $f = \left\{ {{{\tilde \psi }_\mathrm{eff}},{{\tilde F}_r},{{\tilde A}_r}} \right\}$}\\ \noalign{\hrule height 1pt}
		\multicolumn{3}{!{\vrule width 1pt}c!{\vrule width 1pt}}{\bf{continuity of}}\\ \hline 
		$m{{\tilde \psi }_\mathrm{eff}}$ 											& $\mu {{\tilde F}_r}$ 		    & $\varepsilon {{\tilde A}_r}$ \\ 
		${\partial _r}{{\tilde \psi }_\mathrm{eff}}$ 								& ${\partial _r}{{\tilde F}_r}$ & ${\partial _r}{{\tilde A}_r}$ \\ \noalign{\hrule height 1pt}
	\end{tabular}
	\label{table}
\end{table}

Table \ref{table} also reveals that the presented electron--light analogy for spherical waves is very similar to that for plane waves \cite{Jelinek-2011_b, Gaylord-1993, Dragoman-1999, Dragoman-2007}.

\subsection{\label{subsec:B}The embedded eigenstate}
The idea of a trapped electron is inspired by the electromagnetic case \cite{Silveirinha-2014a}, where it was shown that electromagnetic modes can under certain conditions be bound with infinite lifetimes in a core--shell nanoparticle. Particularly, the $\mathrm{TM}^r$ modes can be bound in the inner region of a core--shell nanostructure when the permittivity of the shell is zero--valued, $\varepsilon_\mathrm{shell}=0$, and the radius of the core has a precise value. In such a case, the shell has infinite transverse wave impedance and behaves, for this particular mode of oscillation, as a perfect magnetic conductor (PMC).

Applying the analogy described in the previous section, we see that an electron may be trapped in the core of a spherical heterostructure with an energy such that the effective dispersive mass of the shell vanishes. From Eq. (\ref{eq4}) the condition $m=0$ is satisfied for an energy such that $E=E_{\Gamma_8}^{(2)}$, i.e. at the edge of the valence (with $p$-type symmetry) band. In what follows, we will show that a semiconductor with a zero--valued dispersive mass may, indeed, effectively behave as an infinite barrier for the electron, and enables the emergence of a spatially localized stationary state embedded within the continuum. The geometry of the open quantum resonator is sketched in Fig. \ref{fig1}.

Our design is based on the ternary compound $\mathrm{Hg}_{1-x}\mathrm{Cd}_x\mathrm{Te}$, which offers an opportunity to switch between regular and inverted band structures via a change in the mole fraction $x$ of cadmium \cite{Hansen-1982}. The use of this compound also guarantees an almost lattice matched heterostructure. In our design, both the core ($\mathrm{Hg}_{0.9}\mathrm{Cd}_{0.1}\mathrm{Te}$) and the shell ($\mathrm{HgTe}$) have inverted band structures with the $\Gamma_8$ bands lying above the $\Gamma_6$ bands (Fig. \ref{fig1}), and the core and the background materials are assumed to be identical. The band edge energies are calculated from the width of the band gap $E_g=E_{\Gamma_6}-E_{\Gamma_8}$ and from the split--off energy $\Delta=E_{\Gamma_8}-E_{\Gamma_7}$. Energy $E_g$ is computed from the Hansen's formula \cite{Hansen-1982}, considering zero temperature. The split--off energy is taken as $\Delta=0.93$ eV \cite{Rogalski-2005}. The valence band offset between $\mathrm{HgTe}$ and $\mathrm{Hg}_{1-x}\mathrm{Cd}_x\mathrm{Te}$ (see Fig. \ref{fig1}) is evaluated as $\Lambda=0.35x$ eV \cite{Kowalczyk-1986}. The Kane's parameter $P$ is given by the relation $2P^2m_\mathrm{e}/\hbar^2=18+3x$ eV \cite{Wiley-1969}.
\begin{figure*}[t]
	\includegraphics[width=0.9\textwidth]{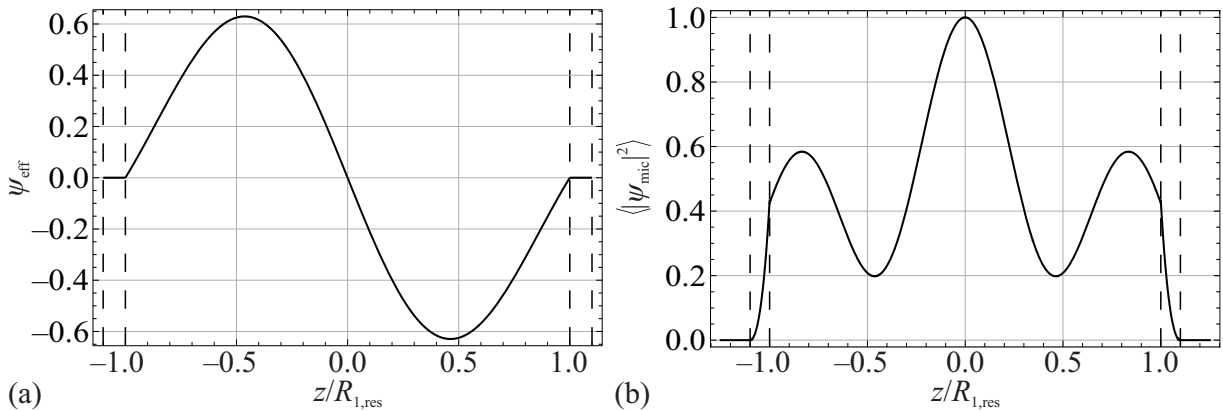}
	\caption{(a) The macroscopic wave function $\psi_\mathrm{eff}$ and (b) the spatial average probability density $\left\langle \left| \psi _\mathrm{mic} \right|^2 \right\rangle $ (normalized to the peak value) of the trapped electron as a function of the normalized radial coordinate.}
	\label{fig2}
\end{figure*}
Next, we formally demonstrate that Eq. (\ref{eq5}) supports a bound state when $E=E_{\Gamma_8}^{(2)}$, i.e. when $m_2=0$ in the shell. Note that the energy level $E=E_{\Gamma_8}^{(2)}$ lies within the continuous energy spectrum of the background and core regions (Fig. \ref{fig1}), so the wave number $k_1=\sqrt{2m_1\left( E-E_{\Gamma_6}^{(1)}\right)/\hbar^2}$ in the core and background regions is real--valued. Under the assumption that $m_2=0$ in the shell, and that the wave function vanishes in the background region, the solution of the radial part of the Schr\"odinger equation (\ref{eq5}) can be written in the form \cite{Arfken}
\begin{equation}
{\tilde R_n}\left( r \right) = {N_0}\left\{ {\begin{array}{*{20}{c}}
	{{j_n}\left( {{k_1}r} \right)}&{r < {R_1}}\\
	{{A_n}{{\left( {{k_1}r} \right)}^n} + {B_n}{{\left( {{k_1}r} \right)}^{ - n - 1}}}&{{R_1} < r < {R_2}}\\
	0&{r > {R_2}}
	\end{array}} \right.
\label{eq7}
\end{equation}
where $N_0$ is a normalization constant. The unknown coefficients $A_n$ and $B_n$ are determined by the boundary conditions, which require the continuity of $m \tilde{R}\left( r \right)$ and $\partial \tilde{R}\left( r \right)$ at the two interfaces. The continuity of $\partial \tilde{R}\left( r \right)$ implies that coefficients $A_n$ and $B_n$ are related as
\begin{equation}
{A_n} = \frac{{{{\left( {{k_1}{R_1}} \right)}^{1 - n}}}}{n}{j'_n}\left( {{k_1}{R_1}} \right){\left[ {1 - {{\left( {\frac{{{R_2}}}{{{R_1}}}} \right)}^{2n + 1}}} \right]^{ - 1}} \nonumber
\end{equation}
and
\begin{equation}
{B_n} = {A_n}\frac{n}{{n + 1}}{\left( {{k_1}{R_2}} \right)^{2n + 1}} \nonumber
\end{equation}
where ${j'_n} \left( x \right) = {\mathrm{d} {j_n} \left( x \right)} /  {\mathrm{d} \left( x \right)}$.

The continuity of $m \tilde{R}\left( r \right)$ imposes that the inner radius must satisfy:
\begin{equation}
j_n\left( k_1R_1\right) = 0
\label{eq8}
\end{equation}
This condition shows that in order to have an embedded energy eigenvalue the radius of the core region must be chosen precisely. For $n=1$ (dipole--type symmetry) this condition implies that the smallest possible radius for the core is $R_{1,\mathrm{res}} \approx 4.49/k_1$. This analysis confirms the hypothesis that the electron can be trapped in the core of the semiconductor heterostructure if the dispersive mass of the shell is zero--valued and the radius of the core has a very specific value.

It is important to highlight that:
\begin{itemize}
	\item In the ideal case of $m_2=0$, the resonance condition is independent of the shell thickness.
	
	\item For $n=0$, the calculated coefficients $A_n$ and $B_n$ are singular, and hence a wave function with monopole (s--type orbital) symmetry cannot be trapped within the core. This important result implies that our resonator is penetrable by electron waves with monopole symmetry, i.e. a semiconductor with a zero--valued dispersive mass behaves as an infinite barrier only for waves with a nonzero azimutal quantum number. Thus, the core--shell heterostructure is generally open to electron waves. This is similar to the electromagnetic case, where the $\mathrm{TM}^r$ wave may be bound to the core by a shell made of permittivity near zero (ENZ) material, with the shell being penetrable by $\mathrm{TE}^r$ waves \cite{Silveirinha-2014a}.
	
	\item The trapped modes are degenerate, because for each $n$ there are in total $2n+1$ spherical harmonics differing only in the magnetic quantum number \cite{Arfken}.
\end{itemize}

To illustrate the proposed theory, we consider a semiconductor heterostructure with an $\mathrm{HgTe}$ shell. The $\mathrm{Hg}_{1-x}\mathrm{Cd}_x\mathrm{Te}$ core has mole fraction $x=0.1$ and radius $R_1 = R_{1,{\mathrm{res}}} \approx 4.49/{k_1} \sim 65a$, where $a=0.65$ nm  is the lattice constant of the considered bulk semiconductor alloys. The radius of the shell is $R_2=1.1R_{1,{\mathrm{res}}}$. The trapped electron state has dipole--type symmetry ($n=1$).

The calculated radius dependence of the ``macroscopic" wave function $\psi_\mathrm{eff}$ and the corresponding averaged probability density $\left\langle {\left| \psi _\mathrm{mic} \right|}^2 \right\rangle$ for $\theta=\pi$ are depicted in Fig. \ref{fig2}. Note that from Eq. (\ref{eq2}) in each layer $\left\langle {\left| \psi _\mathrm{mic} \right|}^2 \right\rangle $ can be written in terms of $\tilde \psi _\mathrm{eff}$ as follows:
\begin{equation}
\left\langle {\left| \psi _\mathrm{mic} \right|}^2 \right\rangle  = m^2\left| {\tilde \psi }_\mathrm{eff} \right|^2 + \hbar ^2\frac{\partial m}{\partial E}\left\| \nabla \tilde \psi _\mathrm{eff} \right\|^2.
\label{eq9}
\end{equation}
To obtain the formula presented above, we used $\partial V_\mathrm{eff}/\partial E = 0$. It is interesting to note that for both semiconductor alloys $m \approx \left( E- E_{\Gamma_8}\right)/{2v_p^2}$ in the energy range of interest, with $v_p^2=2P^2/3\hbar^2$ \cite{Silveirinha-2012a}. The parameter $v_p$ has unities of velocity. Thus $\partial m / \partial E \approx 1/2v_p^2$, which is approximately the same in both the core and the shell. 

Figure \ref{fig2}a shows that the ``macroscopic" electron wave function is entirely confined within the core, i.e. $\psi_\mathrm{eff}$ is identically zero not only outside the core--shell resonator, but also in the shell itself. However, as is shown in Fig. \ref{fig2}b, the average probability density is nonzero in the shell. This means that the microscopic wave function ($\psi_\mathrm{mic}$) has strong fluctuations on the scale of the unit cell of the $\mathrm{HgTe}$ shell, so that its macroscopic spatial average vanishes in the shell, while the corresponding probability density function is nonzero. The fact that the probability of finding the electron in the shell is nonzero is consistent with the electromagnetic case, for which the electromagnetic energy stored in the permittivity near zero (ENZ) shell is nonzero. Thus, a zero dispersive mass, $m_2=0$, and a nonzero azimutal quantum number, imply that the shell behaves as an infinite height potential barrier that blocks the electron tunneling out of the resonator.

It is relevant to note that in the electromagnetic case the light remains confined in the core region due to the screening provided by the (nonradiative) volume plasmons of the shell \cite{Silveirinha-2014a}. Interestingly, in the semiconductor case studies here the role of the plasmons is played by the heavy--hole states of HgTe \cite{Silveirinha-2014b}. In our framework the heavy--hole states have a flat energy dispersion and occur precisely at the energy level wherein the dispersive mass vanishes.
\section{\label{sec:level3}The Trapping Lifetime for a Detuned Resonator}
The previous section dealt with the ideal case, where the energy of the trapped electron is equal to the band edge energy - $E_{\Gamma_8}$ - of the material in the shell, and the inner radius is perfectly tuned to the value $R_{1,\mathrm{res}}$ defined by Eq. (\ref{eq8}). Such perfect tuning is however unrealistic, and it is interesting to characterize the trapping lifetime when the inner radius $R_1$ is detuned.
\begin{figure}[!ht]
	\includegraphics[width = 0.9\columnwidth]{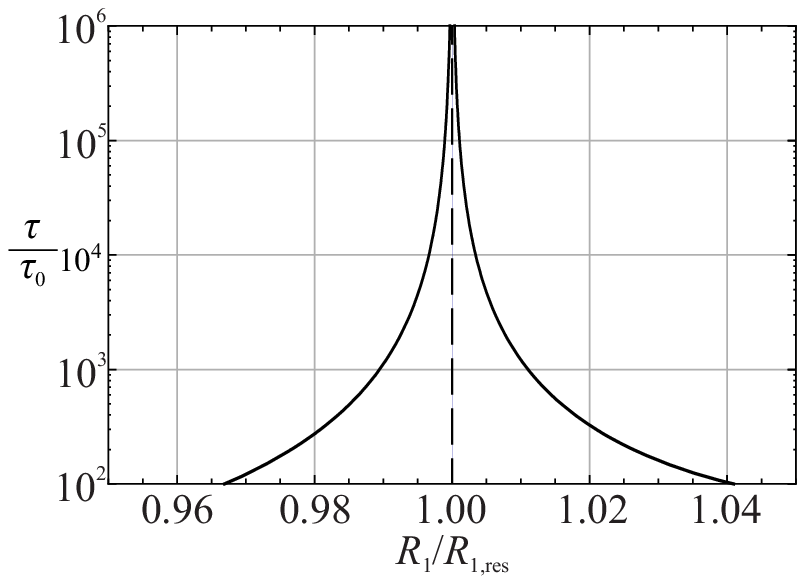}
	\caption{The trapping lifetime as a function of relative detuning $R_1 / R_{1,\mathrm{res}}$. The trapping lifetime is normalized with respect to the time $\tau_0 = 2 R_2 / v_g = 0.11$ ps that the electron needs to pass the diameter of the core--shell resonator at the group velocity \small{$ v_g = v_p \sqrt{\left( E - E_{\Gamma_6}^{(3)} \right)\left( E - E_{\Gamma_8}^{(3)} \right)} / \left( E - \left( E_{\Gamma_6}^{(3)} - E_{\Gamma_8}^{(3)} \right) / 2 \right)$} in the background material.}
	\label{fig3}
\end{figure}

In the detuned case, the solution of the radial equation (\ref{eq5}) has to be searched in the form
\begin{equation}
\tilde R\left( r \right) = \left\{ {\begin{array}{*{20}{c}}\displaystyle
	{{a_n}{j_n}\left( {{k_1}r} \right)}&{r < {R_1}}\\
	{b_n^{(1)}{j_n}\left( {{k_2}r} \right) + b_n^{(2)}{y_n}\left( {{k_2}r} \right)}&{{R_1} < r < {R_2}}\\
	{{c_n}h_n^{\left( 1 \right)}\left( {{k_3}r} \right)}&{r > {R_2}}
	\end{array}} \right.
\label{eq10}
\end{equation}
where $j_n$, $y_n$ are the spherical Bessel functions of the first and second kind, respectively, $h_n^{(1)}$ is the spherical Hankel function of the first kind and $k_i = \sqrt{2m_i \left( E-E_{\Gamma_6}^{(i)} \right) / \hbar^2}$ is the wave number in the $i$-th layer. As in the previous section, the unknown coefficients $a_n$, $b_n^{(1)}$, $b_n^{(2)}$ and $c_n$ are obtained from the boundary conditions discussed previously, which result in the following equation system 
\begin{widetext}
	\begin{equation}
	\left( {\begin{array}{*{20}{c}}
		{{j_n}\left( {{k_1}{R_1}} \right)}    & { - \displaystyle\frac{{{m_2}}}{{{m_1}}}{j_n}\left( {{k_2}{R_1}} \right)}    & { - \displaystyle\frac{{{m_2}}}{{{m_1}}}{y_n}\left( {{k_2}{R_1}} \right)}    & 0\\
		{{{j'}_n}\left( {{k_1}{R_1}} \right)} & { - \displaystyle\frac{{{k_2}}}{{{k_1}}}{{j'}_n}\left( {{k_2}{R_1}} \right)} & { - \displaystyle\frac{{{k_2}}}{{{k_1}}}{{y'}_n}\left( {{k_2}{R_1}} \right)} & 0\\
		0 									  & {\displaystyle\frac{{{m_2}}}{{{m_1}}}{j_n}\left( {{k_2}{R_2}} \right)}       & \displaystyle{\frac{{{m_2}}}{{{m_1}}}{y_n}\left( {{k_2}{R_2}} \right)}       & { - \displaystyle\frac{{{m_3}}}{{{m_1}}}h_n^{\left( 1 \right)}\left( {{k_3}{R_2}} \right)}\\
		0									  & {\displaystyle\frac{{{k_2}}}{{{k_1}}}{{j'}_n}\left( {{k_2}{R_2}} \right)}    & {\displaystyle\frac{k_2}{k_1}{{y'}_n}\left( {k_2}{R_2} \right)}    & { - \displaystyle\frac{k_3}{k_1} {h'}_n^{\left( 1 \right)} \left( {k_3}{R_2} \right)}
		\end{array}} \right)\left( {\begin{array}{*{20}{c}}
		{{a_n}}\\
		{b_n^{(1)}}\\
		{b_n^{(2)}}\\
		{{c_n}}
		\end{array}} \right) = 0
	\label{eq11}
	\end{equation}
\end{widetext}

In the detuned case, this homogeneous system (\ref{eq10}) has a non-trivial solution only for complex energy values, $E=E_\mathrm{re}+iE_\mathrm{im}$, which correspond to the zeros of the matrix determinant. 
The imaginary part of the energy is associated with the decay time of the localized state, and nonzero $E_\mathrm{im}$  implies that the electron escapes from the resonator. The trapping lifetime can be defined as $\tau  \sim \hbar /\left( { - 2{E_{{\mathrm{im}}}}} \right)$ \cite{Gottfried-2003}. The lifetime is independent of the origin of the energy scale. The trapping lifetime is shown in Fig. \ref{fig3} as a function of relative detuning $R_1 / R_{1,\mathrm{res}}$ for $R_2 = 1.1R_{1,\mathrm{res}}$. The calculation assumes that the core and the background are made of ${\mathrm{Hg}_{0.9}\mathrm{Cd}_{0.1}\mathrm{Te}}$, and that the shell is $\mathrm{HgTe}$.

\begin{figure*}[ht]
	\includegraphics[width=0.9\textwidth]{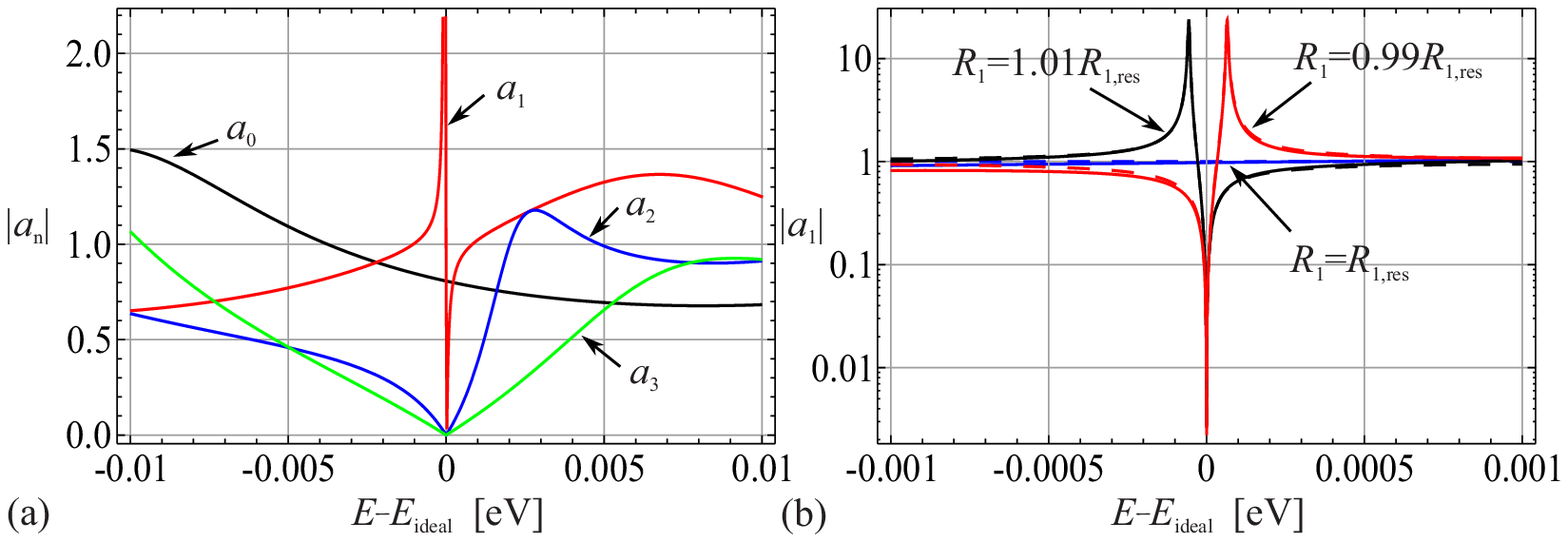}
	\caption{(color online) (a)Magnitude of the first four coefficients $a_n$ ($n = 0, 1, 2, 3$) as a function of relative energy detuning. The inner radius of the resonator is ${R_1} = 1.01R_{1,\mathrm{res}}$. (b) Magnitude of $a_1$ as a function of relative energy detuning. The inner radius is ${R_1} = \left\lbrace 0.99,1,1.01\right\rbrace R_{1,\mathrm{res}}$. The full lines correspond to the exact solution while the dashed lines correspond to the approximation (\ref{eq14}).}
	\label{fig4}
\end{figure*}
\section{\label{sec:level4}Scattering Cross--section of the Core--shell Resonator}
Since the resonator may support a state with an infinite lifetime, it is natural to ask if it can capture a free--electron propagating in the background region. To investigate this possibility, we will now study the scattering of a plane electron wave by the core--shell resonator.
	
Because of the angular symmetry of the resonator, it can be assumed without loss of generality that the plane wave propagates along the $z$-axis. This plane wave may be decomposed into Legendre polynomials as \cite{Jackson01}
\begin{equation}
\mathrm{e}^{\mathrm{i}{k_3}z} = \sum\limits_{n = 0}^\infty  \mathrm{i}^n\left( 2n + 1 \right)j_n\left( kr \right)P_n\left( {\cos \theta } \right) 
\label{eq12}
\end{equation}
This decomposition allows us to write the normalized wave function as $\tilde{\psi}= \sum_{0}^{\infty}\tilde{\psi_n}$ with ${\tilde \psi}_n\left( {r,\theta ,\varphi } \right) = {\tilde R_n}\left( r \right){\mathrm{i}^n} \left( {2n + 1} \right) {P_n}\left(\cos{ \theta} \right)$ and
\begin{equation}
{\tilde R_n}\left( r \right) = \left\{ {\begin{array}{*{20}{c}}
{{a_n}{j_n}\left( {{k_1}r} \right)}&{r < {R_1}}\\
{b_n^{(1)}{j_n}\left( {{k_2}r} \right) + b_n^{(2)}{y_n}\left( {{k_2}r} \right)}&{{R_1} < r < {R_2}}\\
{{c_n}h_n^{\left( 1 \right)}\left( {{k_3}r} \right) + {j_n}\left( {{k_3}r} \right)}&{r > {R_2}}
\end{array}} \right.
\label{eq13}
\end{equation}
\begin{figure*}[ht]
	\includegraphics[width = 0.9\textwidth]{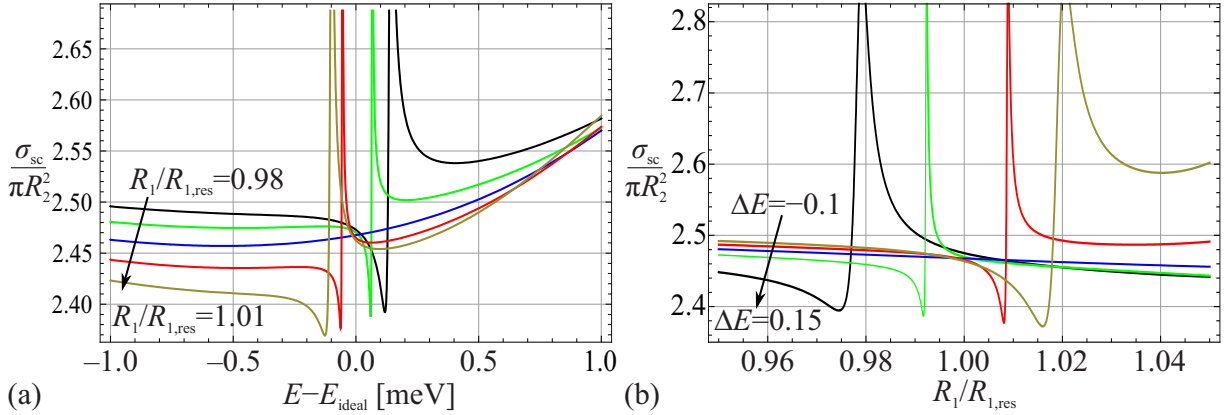}
	\caption{(color online) (a) Scattering cross section $\sigma_\mathrm{sc}$ as a function of relative energy detuning for ${R_1} / R_{1,\mathrm{res}}=0.98, 0.99, 1, 1.01$ and $1.02$. (b) Scattering cross section $\sigma_{\mathrm{sc}}$  as a function of the inner radius ${R_1} / R_{1,\mathrm{res}}$ for incident electron energy $E=E_{\mathrm{ideal}}+\Delta E$, where $\Delta E = \left\lbrace 0.15, 0.1, 0, -0.05, -0.1 \right\rbrace $ meV. In all calculation outer radius was taken as $R_2 = 1.1R_{1,\mathrm{res}}$.}	
	\label{fig5}
\end{figure*}
The unknown coefficients are obtained by imposing the previously discussed boundary conditions at the interfaces. Figure \ref{fig4}a shows the first four ($n= 0, 1, 2, 3$) Mie scattering coefficients in the core region ($a_n$) as a function of the electron energy for a detuned resonator with $R_1 = 1.01 R_{1,\mathrm{res}}$. The curve corresponding to $n=0$ (black line) confirms that the heterostructure is penetrable by an electron wave with monopole symmetry. This is in conformity with the claim in Sec. \ref{subsec:B}. However, the remaining curves show clearly that $a_n$ ($n \ge 1$) vanishes whenever $E = E_\mathrm{ideal}$ (see Fig. \ref{fig4}a), where $E_\mathrm{ideal}$ is the valence band edge energy $E_{\Gamma_8}$ of HgTe. For this energy, the shell region behaves as an infinite height barrier, and the incident electron wave is unable to reach the core region. The most relevant of these coefficients, $a_1$, which is associated with the trapped state, is further studied in Fig. \ref{fig4}b for ${R_1} \approx R_{1,\mathrm{res}}$. In this case, our numerical simulations reveal that the approximation
\begin{equation}
a_1 \approx \frac{\left( E-E_\mathrm{ideal}\right)}{\left( E-E_\mathrm{actual}\right)}\mathrm{e}^{i\phi_0}
\label{eq14}
\end{equation}
holds. In the above, $E_\mathrm{actual}$ is the complex valued resonance energy determined by the inner radius $R_1$ and which is calculated as is explained in Sect. \ref{sec:level3}, and $\phi_0$ is some irrelevant phase factor. Notably, equation (\ref{eq14}) and Fig. \ref{fig4}b reveal that in the limit case $R_1 \to R_{1,\mathrm{res}}$ the zero associated with $E_\mathrm{ideal}$ cancels the pole corresponding to $E_\mathrm{actual}$, and $\left| a_1 \right| \to 1$. This contrasts with all the other $a_n$ ($n \ge 2$), which in the present example vanish identically for $E=E_\mathrm{ideal}$, regardless of the radius $R_1$. This means that, due to the cancellation of a zero--pole, an incident wave with energy $E=E_\mathrm{ideal}$ and dipole--symmetry $may$ actually penetrate into the shell, in the case of a perfectly tuned resonator (see Fig. \ref{fig4}b, blue curve). Nevertheless, even though the resonator may support an infinite lifetime bound state and the free electron can penetrate into the core, it cannot be captured by the resonator. 

Indeed, the condition for having a trapped electron in the present problem is that $\left| a_1 \right| \to \infty$ for some real--valued $E$. It may be checked that even though $\left| a_1 \right|$ can have rather large values in our structure, it remains finite for any real-valued energy. We therefore conclude that, in the scenario studied here, the resonator is unable to capture the free electron.
	
This discussion may suggest that it is impossible to couple a free electron to the embedded bound state. However, that is not necessarily the case. For example, if the resonator is perturbed during a short time period (e.g. by applying a time-varying electric or magnetic field), the temporary detuning may allow the free electron to excite the bound state and be permanently captured after the perturbation is removed. These ideas will be explored in future work.

It is also interesting to characterize the scattering cross--section of the resonator. It is given by \cite{Morse}
\begin{widetext}	
\begin{equation}
{\sigma _{{\mathrm{sc}}}} = \int\limits_\Omega  {\frac{{\left| {{\psi _{{\mathrm{sc}}}}\frac{{\partial \psi _{{\mathrm{sc}}}^ * }}{{\partial r}} - \psi _{{\mathrm{sc}}}^ * \frac{{\partial {\psi _{{\mathrm{sc}}}}}}{{\partial r}}} \right|{r^2}}}{{\left| {{\psi _{{\mathrm{inc}}}}\nabla \psi _{{\mathrm{inc}}}^ *  - \psi _{{\mathrm{inc}}}^ * \nabla {\psi _{{\mathrm{inc}}}}} \right|}}{\mathrm{d}}\Omega }  = \frac{{4\pi }}{{k_1^2}}\sum\limits_n {{{\left| {{c_n}} \right|}^2}\left( {2n + 1} \right)} 
\label{eq15}
\end{equation}
\end{widetext}
where subscript ``sc" stands for scattered, and subscript ``inc" stands for incident. Figure \ref{fig5}a shows that for a perfectly tuned resonator with $R_1=R_{1,\mathrm{res}}$ (blue curve), the scattering cross section does not exhibit any resonant features. This is consistent with the zero--pole cancellation discussed above. However, for a detuned inner radius $R_1$ there is a resonant response which indicates a strong interaction of the free--electron with the heterostructure because of temporary electron trapping. This behavior is also perceptible in Fig. \ref{fig5}b, where the scattering cross--section is represented for different energies of the incident electron.
\section{Conclusion}

It has been shown that a spherical semiconductor heterostructure may support bound states embedded within the continuum at an energy level where the shell region has a zero--valued dispersive mass. A realistic design of the heterostructure based on the $\mathrm{Hg}_{1-x}\mathrm{Cd}_x\mathrm{Te}$ compound has been proposed. An in-depth analysis of the suggested heterostructure has been presented, showing the possibility to trap an electron within the resonator core. The trapping lifetime of a detuned heterostructure has also been characterized, and it has been shown that the heterostructure can trap the electron for a long time,  even if there is slight detuning. Finally, we investigated the possibility of a free electron being captured by the semiconductor resonator. Notably, our analysis has revealed that, in the same manner as a trapped electron is unable to escape from the resonator, a free electron cannot be permanently captured by the resonator. Interestingly, the scattering cross section does not exhibit any resonant features for a perfectly tuned structure. This confirms that a free electron is unable to interact with the embedded bound energy eigenstate.
Although it is challenging to capture the incident electron within the core of an ideal structure, our future aim will be to investigate how to couple the electron to the embedded bound state, e.g. by a temporary detuning of the heterostructure.
\section*{Appendix A}
In the supplementary materials of \cite{Fernandes-2014} it is formally demonstrated that, within an effective medium framework, the relation between the spatially averaged probability density associated with a Bloch energy eigenstate and the macroscopic wave function is such that:
\begin{equation}\tag{A1}
\left\langle {{{\left| {{\psi _{{\mathrm{mic}}}}} \right|}^2}} \right\rangle  = \left( {1 - \frac{{\partial {{\hat H}_{{\mathrm{eff}}}}}}{{\partial E}}} \right){\left| {{\psi _{\mathrm{eff}}}} \right|^2}
\label{app1}
\end{equation}
In the above, ${\hat H_{{\mathrm{eff}}}}\left( {E,{\bf{k}}} \right)$  represents the homogenized (energy--dependent) Hamiltonian with $\mathrm{\bf{k}}=-i\nabla$. In this paper, the effective medium Hamiltonian is given by [see Eq. (\ref{eq3})]:
\begin{equation}\tag{A2}
{\hat H_{{\mathrm{eff}}}}\left( {E,{\bf{k}}} \right) = \frac{{{\hbar ^2}{k^2}}}{{2m}} + {V_{{\mathrm{eff}}}}
\label{app2}
\end{equation}
Hence, it follows that:
\begin{equation}\tag{A3}
\left\langle {{{\left| {{\psi _{{\mathrm{mic}}}}} \right|}^2}} \right\rangle  = \left( {1 - \frac{{\partial {V_{{\mathrm{eff}}}}}}{{\partial E}}} \right){\left| {{\psi _{\mathrm{eff}}}} \right|^2} - \frac{\partial }{{\partial E}}\left( {\frac{1}{m}} \right)\frac{{{\hbar ^2}}}{2}{\left| {i{\mathrm{\bf{k}}}{\psi _{\mathrm{eff}}}} \right|^2}
\label{app3}
\end{equation}
For a Bloch energy eigenstate in a continuous medium, we have $i{\bf{k}}{\psi _{\mathrm{c}}} = \nabla {\psi _{\mathrm{c}}}$, and thus the above result leads to Eq. (\ref{eq2}) of the main text.
\section*{Acknowledgment}
This work has been supported by the Czech Science Foundation under project No. 13-09086S, and by Funda\c{c}\~ao para a Ci\^encia e Tecnologia, under project PTDC/EEI-TEL/2764/2012.

\bibliography{References2013}

\begin{thebibliography}{10}

\bibitem{Gottfried-2003}
K.~Gottfried and T.-M. Yan.
\newblock {\em Quantum mechanics: Fundamentals}, page 620.
\newblock Springer, 2nd edition, 2003.

\bibitem{Neumann-1929}
J.~von Neumann and E.~Wigner.
\newblock {\em Phys. Z.}, 30:467, 1929.

\bibitem{Stillinger-1975}
F.~H. Stillinger and D.~R. Herrick.
\newblock Coherent potential approximation: Basic concepts and applications.
\newblock {\em Phys. Rev. A}, 11:446, 1975.

\bibitem{Feshbach-1958}
H.~Feshbach.
\newblock Unified theory of nuclear reactions.
\newblock {\em Annals of Physics (N.Y.)}, 5:357, 1958.

\bibitem{Newton}
R.~G. Newton.
\newblock {\em Scattering Theory of Waves and Particles}.
\newblock Springer-Verlag, 2nd edition, 2013.

\bibitem{Bulgakov-2006}
E.~N. Bulgakov, K.~N. Pichugin, and I.~Rotter.
\newblock Bound states in the continuum in open aharonov--bohm ring.
\newblock {\em JETP Letters}, 84:430--435, 2006.

\bibitem{Sadreev-2006}
A~F. Sadreev, E.~N. Bulgakov, and I.~Rotter.
\newblock Bound states in the continuum in open quantum billiards with a
  variable shape.
\newblock {\em Phys. Rev. B}, 73:23534, 2006.

\bibitem{Moiseyev-2009}
N.~Moiseyev.
\newblock Suppression of feshbach resonance widths in two-dimensional
  waveguides and quantum dots: A lower bound for the number of bound states in
  the continuum.
\newblock {\em Phys. Rev Lett.}, 102:167404, 2009.

\bibitem{Capasso-1992}
F.~Capasso, C.~Sirtori, J.~Faist, D.~L. Sivco, S.-N.~G. Chu, and A.~Y. Cho.
\newblock Observation of an electronic bound states above a potential well.
\newblock {\em Nature}, 358:565, 1992.

\bibitem{Weber-1994}
T.~A. Weber.
\newblock Bound states with no classical turning points in semiconductor
  heterostructures.
\newblock {\em Solid State Commun.}, 90:713, 1994.

\bibitem{Gippius-2005}
N.~A. Gippius, S.~G. Tikhodeev, and T.~Ishihara.
\newblock Optical properties of photonic crystal slabs with an asymmetrical
  unit cell.
\newblock {\em Phys. Rev. B}, 72:045138, 2005.

\bibitem{Borisov-2005}
A.~G. Borisov, F.~J.~García de~Abajo, and S.~V. Shabanov.
\newblock Role of electromagnetic trapped modes in extraordinary transmission
  in nanostructured materials.
\newblock {\em Phys. Rev. B}, 71:075408, 2005.

\bibitem{Marinica-2008}
D.~C. Marinica, A.~G. Borisov, and S.~V. Snabanov.
\newblock Bound states in the continuum in photonics.
\newblock {\em Phys. Rev. Lett.}, 100:183902, 2008.

\bibitem{Silveirinha-2014a}
M.~G. Silveirinha.
\newblock Trapping light in open plasmonic nanostructures.
\newblock {\em Phys. Rev. A}, 89:023813, 2014.

\bibitem{Alu-2014}
F.~Monticone and A.~Alù.
\newblock Embedded photonic eigenvalues in 3d nanostructures.
\newblock {\em Phys. Rev. Lett.}, 112:213903, 2014.

\bibitem{Plotnik-2011}
Y.~Plotnik, O.~Peleg, F.~Dreisow, M.~Heinrich, S.~Nolte, A.~Szameit, and
  M.~Segev.
\newblock Experimental observation of optical bound states in the continuum.
\newblock {\em Phys. Rev. Lett.}, 107:183901, 2011.

\bibitem{Lee-2012}
J.~Lee, B.~Zhen, S.-L. Chua, W.~Qiu, J.~D. Joannopoulos, M.~Soljacic, and
  O.~Shapira.
\newblock Observation and differentiation of unique high-q optical resonances
  near zerowave vector in macroscopic photonic crystal slabs.
\newblock {\em Phys. Rev. Lett.}, 109:067401, 2012.

\bibitem{Hsu-2013}
Ch.~W. Hsu1, B.~Zhen, J.~Lee, S.-L. Chua, S.~G. Johnson, J.~D. Joannopoulos,
  and M.~Soljacic.
\newblock Observation of trapped light within the radiation continuum.
\newblock {\em Nature}, 499:188, 2013.

\bibitem{Harald}
Harald~Siegfried Friedrich.
\newblock {\em Theoretical Atomic Physics}, page 506.
\newblock Springer Berlin Heidelberg, 2006.

\bibitem{Kane-1957}
E.~O. Kane.
\newblock Band structure of indium antimonide.
\newblock {\em J. Phys. Chem. Sol.}, 1:249, 1957.

\bibitem{Bastard-1986}
G.~Bastard and J.~A. Brum.
\newblock Electronic states in semiconductor heterostructures.
\newblock {\em IEEE J. Quantum Elect.}, 22:1625, 1986.

\bibitem{Silveirinha-2012a}
M.~G. Silveirinha and N.~Engheta.
\newblock Effective medium approach to electron waves: Graphene superlattices.
\newblock {\em Phys. Rev. B}, 85:195413, 2012.

\bibitem{Lannebere-2015}
S.~Lanneb\`ere and M.~G. Silveirinha.
\newblock Effective hamiltonian for electron waves in artificial graphene: A
  first-principles derivation.
\newblock {\em Phys. Rev. B}, 91:045416, 2015.

\bibitem{Silveirinha-2012c}
M.~G. Silveirinha and N.~Engheta.
\newblock Transformation electronics: Tailoring the effective mass of
  electrons.
\newblock {\em Phys. Rev. B}, 86:161104, 2012.

\bibitem{Silveirinha-2012b}
M.~G. Silveirinha and N.~Engheta.
\newblock Metamaterial-inspired model for electron waves in bulk
  semiconductors.
\newblock {\em Phys. Rev. B}, 86:245302, 2012.

\bibitem{Bastard}
G.~Bastard.
\newblock {\em Wave Mechanics Applied to Semiconductor Heterostructures}.
\newblock John Wiley \& Sons, 1988.

\bibitem{Bastard-1982}
G.~Bastard.
\newblock Theoretical investigations of superlattice band structure in the
  envelope-function approximation.
\newblock {\em Phys. Rev. B}, 25:7584, 1982.

\bibitem{Fernandes-2014}
D.~E. Fernandes, N.~Engheta, and M.~G. Silveirinha.
\newblock Wormhole for electron waves in graphene.
\newblock {\em Phys. Rev. B}, 90:041406(R), 2014.

\bibitem{Arfken}
G.B. Arfken and H.J. Weber.
\newblock {\em Mathematical Methods for Physicists}, page 981.
\newblock Harcourt/Academic Press, 5th edition, 2001.

\bibitem{Balanis}
C.~A. Balanis.
\newblock {\em Antenna Theory: Analysis and Design}, page 950.
\newblock John Wiley \& Sons, Inc., 2nd edition, 1997.

\bibitem{Jelinek-2011_b}
L.~Jelinek, J.~D. Baena, J.~Voves, and R.~Marques.
\newblock Metamaterial-inspired perfect tunnelling in semiconductor
  heterostructures.
\newblock {\em New J. Phys.}, 13:083011, 2011.

\bibitem{Gaylord-1993}
T.~K. Gaylord, G.~N. Henderson, and E.~N. Glytsis.
\newblock Application of electromagnetics formalism to quantum-mechanical
  electron-wave propagation in semiconductors.
\newblock {\em J. Opt. Soc. Am. B}, 10:333, 1993.

\bibitem{Dragoman-1999}
D.~Dragoman and M.~Dragoman.
\newblock Optical analogue structures to mesoscopic devices.
\newblock {\em Prog. Quant. Electron.}, 23:131, 1999.

\bibitem{Dragoman-2007}
D.~Dragoman and M.~Dragoman.
\newblock Metamaterials for ballistic electrons.
\newblock {\em J. Appl. Phys.}, 101:104316, 2007.

\bibitem{Hansen-1982}
J.~L.~Schmitt G.~L.~Hansen and T.~N. Casselman.
\newblock Energy gap versus alloy composition and temperature in hg1−x cd x
  te.
\newblock {\em J. Appl. Phys.}, 53:7099, 1982.

\bibitem{Rogalski-2005}
A.~Rogalski.
\newblock Hgcdte infrared detector material: history, status and outlook.
\newblock {\em Rep. Prog. Phys.}, 68:2267, 2005.

\bibitem{Kowalczyk-1986}
S.~P. Kowalczyk, J.~T. Cheung, E.~A. Kraut, and R.~W. Grant.
\newblock Cdte-hgte (111) heterojunction valence-band discontinuity: A
  common-anion-rule contradiction.
\newblock {\em Phys. Rev. Lett.}, 56:1605, 1986.

\bibitem{Wiley-1969}
John~D. Wiley and R.~N. Dexter.
\newblock Helicons and nonresonant cyclotron absorption in semiconductors. ii.
  ${\mathrm{hg}}_{1-x}{\mathrm{cd}}_{x}\mathrm{Te}$.
\newblock {\em Phys. Rev.}, 181:1181--1190, May 1969.

\bibitem{Silveirinha-2014b}
M.~G. Silveirinha and N.~Engheta.
\newblock Giant nonlinearity in zero-gap semiconductor superlattices.
\newblock {\em Phys. Rev. B}, 89:085205, 2014.

\bibitem{Jackson01}
J.~D. Jackson.
\newblock {\em Classical Electrodynamics}.
\newblock John Wiley \& Sons, Inc., 3rd edition, 1998.

\bibitem{Morse}
P.~M. Morse and H.~Feshbach.
\newblock {\em Methods of Theoretical Physics}.
\newblock McGraw-Hill Book Company, Inc., 1953.

\end{thebibliography}
\bibliographystyle{unsrt}

\end{document}